\documentclass{article}
\usepackage{eurosym}
\usepackage{amsfonts}
\usepackage{amsmath}

\newtheorem{theorem}{Theorem}
\newtheorem{acknowledgement}[theorem]{Acknowledgement}

\input{tcilatex}
\begin{document}

\title{Einstein-Chern-Simons equations on the 3-brane world}
\author{F. Izaurieta$^{1}$, P. Salgado$^{2}$ and R. Salgado$^{1}$ \\
$^{1}$Departamento de F\'{\i}sica, Universidad de Concepci\'{o}n\\
Casilla 160-C, Concepci\'{o}n, Chile\\
$^{2}$Instituto de Ciencias Exactas y Naturales (ICEN)\\
Facultad de Ciencias, Universidad Arturo Prat\\
Avda. Arturo Prat 2120, Iquique, Chile}
\maketitle

\begin{abstract}
In this article it is studied the 3-brane world in the context of
five-dimensional Einstein-Chern-Simons gravity. We started by considering
Israel's junction condition for AdS-Chern-Simons gravity. Using the $S$%
-expansion procedure, we mapped the AdS-Chern-Simons junction conditions to
Einstein-Chern-Simons gravity, allowing us to derive effective
four-dimensional Einstein-Chern-Simons field equations.
\end{abstract}

\section{\textbf{Introduction}}

The observations and experiments show that General Relativity and the
Standard Model provide the current understanding of the natural phenomena.
From a theoretical point of view, however, the Standard Model are gauge
theories, i.e., they are theories whose fundamental field, is a connection,
while General Relativity is a theory whose fundamental field is a metric.

A gauge theory for the gravitational field requires a fundamental field
given by a connection. An action for gravity fulfilling these conditions is
the Chern-Simons gravity action, which was proposed long ago by Chamseddine 
\cite{cham1},\cite{cham2},\cite{cham3}.

This Chern-Simons gravity is a well-defined gauge theory, but the presence
of higher powers of the curvature makes its dynamics very remote from that
for standard Einstein Hilbert gravity. \ However in Refs. \ \cite{salg1}, 
\cite{edel}, \cite{concha}\textbf{\ }was shown that the standard,
five-di\-men\-sio\-nal General Relativity can be obtained from Chern-Simons
gravity theory for the Lie algebra $\mathfrak{B}_{5}$, whose generators $%
\left\{ J_{AB},P_{A},Z_{AB},Z_{A}\right\} $ satisfy the commutation
relationships

\begin{equation*}
\left[ J_{AB},J_{CD}\right] =\eta _{CB}J_{AD}-\eta _{CA}J_{BD}+\eta
_{DB}J_{CA}-\eta _{DA}J_{CB}
\end{equation*}%
\begin{equation*}
\left[ J_{AB},P_{C}\right] =\eta _{CB}P_{A}-\eta _{CA}P_{B}
\end{equation*}%
\begin{equation*}
\text{\ }\left[ P_{A},P_{B}\right] =Z_{AB}
\end{equation*}%
\begin{equation*}
\left[ J_{AB},Z_{CD}\right] =\eta _{CB}Z_{AD}-\eta _{CA}Z_{BD}+\eta
_{DB}Z_{CA}-\eta _{DA}Z_{CB}
\end{equation*}%
\begin{equation}
\left[ J_{AB},Z_{C}\right] =\eta _{CB}Z_{A}-\eta _{CA}Z_{B}  \label{cuatro}
\end{equation}%
\begin{equation*}
\left[ Z_{AB},P_{C}\right] =\eta _{CB}Z_{A}-\eta _{CA}Z_{B},
\end{equation*}%
which can be obtained from the AdS algebra by means of the S-expansion
procedure introduced in Refs. \cite{salg2}, \cite{salg3}, \cite{saka}, \cite%
{azcarr}. \ An expansion is, in general, an algebra dimension-changing
process, i.e., is a way to obtain new algebras of increasingly higher
dimensions from a given one. \ A physical motivation for increasing the
dimension of Lie algebras is that increasing the number of generators of an
algebra is a non-trivial way of enlarging spacetime symmetries. Examples of
this can be found in Refs. \cite{bonanos}, \cite{gomis}, where applications
of Maxwell's algebra in gravity were studied (This algebra is a modification
to the Poincar\'{e} symmetries and can be obtained, via S-expansion, from
the anti-de Sitter (AdS) which is also known as $\mathfrak{B}_{4}$ algebra).
Another interesting modification to the Poincar\'{e} symmetries are the
so-called generalized Poincar\'{e} algebras \cite{seba} of which the $%
\mathfrak{B}_{5}$ algebra is an example.

For this reason the Chern-Simons gravity theory for the $\mathfrak{B}_{5}$
Lie algebra, known as Einstein-Chern-Simons gravity \cite{coment2} can be
understood as a theory that could allow us to know that if the space-time
has (or not) more symmetry than those usually described by Poincar\'{e} or
(A)dS algebras. This can be achieved by studying cosmological and black hole
solutions such as that found in Refs. \cite%
{uno,dos,tres,cuatro,cinco,seis,siete}. In particular in \cite{siete} EChS
gravity was considered instead of General Relativity to describe the
expansion of a flat 5-dimensional universe, where a cosmological analysis
was performed. The $k^{ab}$-field was assumed null by virtue of the gauge
freedom and the $h^{a}$ field was represented as a perfect fluid. Was found
an accelerating Dirac-Milne universe and a fluid, that early behaves like
dark matter but later behaves like stiff matter. In the same reference was
curry out a compactification $5D$ to $4D$ and was found that in a 4D
context, it is possible to conjecture the existence of accelerated
solutions, eventually driven by the $h$-field. The $h$-field was associated
with a scalar field, exhibiting the behavior of cosmological constant (dark
energy).

In order to write down a Chern--Simons lagrangian for the $\mathfrak{B}_{5}$
algebra, we start from the one-form gauge connection%
\begin{equation}
\boldsymbol{A}=\frac{1}{2}\omega ^{AB}\boldsymbol{J}_{AB}+\frac{1}{l}e^{A}%
\boldsymbol{P}_{A}+\frac{1}{2}k^{AB}\boldsymbol{Z}_{AB}+\frac{1}{l}h^{A}%
\boldsymbol{Z}_{A},  \label{cinco}
\end{equation}%
and the two-form curvature%
\begin{eqnarray}
\boldsymbol{F} &=&\frac{1}{2}R^{AB}\boldsymbol{J}_{AB}+\frac{1}{l}T^{A}%
\boldsymbol{P}_{A}+\frac{1}{2}\left( \mathrm{D}_{\omega }k^{AB}+\frac{1}{%
l^{2}}e^{A}e^{B}\right) \boldsymbol{Z}_{AB}  \notag \\
&&+\frac{1}{l}\left( \mathrm{D}_{\omega }h^{A}+k_{\text{ \ \ }%
B}^{A}e^{B}\right) \boldsymbol{Z}_{A}.  \label{seis}
\end{eqnarray}

In this point, it might be of interest to remember that: $\left( i\right) $
clearly $l$ could be eliminated by absorbing it in the definition of the
vielbein, but then the space-time metric $g_{\mu \nu }$ would no longer be
related to $e^{a}$ through the relation $g_{\mu \nu }=\eta _{ab}e_{\mu }^{%
\text{ }a}e_{\nu }^{\text{ }b}$; $\left( ii\right) $ the interpretation of
the $l$ parameter as a parameter related to the radius of curvature of the $%
AdS$ space-time, could be inherited for the space-time whose symmetries are
described by the $\mathfrak{B}_{5}$ generalized Poincar\'{e} algebra. This
can be seen by recalling that the commutation relation $\left[ P_{a},P_{b}%
\right] =\frac{1}{l^{2}}Z_{ab}$ is obtained, using the expansion method,
from the commutation relation $\left[ \tilde{P}_{a},\tilde{P}_{b}\right] =%
\frac{1}{l^{2}}J_{ab}$ for $AdS$ translations, 
\begin{equation*}
\left[ P_{a},P_{b}\right] =\lambda _{2}\left[ \tilde{P}_{a},\tilde{P}_{b}%
\right] =\frac{1}{l^{2}}\lambda _{2}J_{ab}=\frac{1}{l^{2}}Z_{ab}
\end{equation*}%
where $\lambda _{2}$ is an element of the semigroup $S_{E}^{(3)}$.

In Ref. \cite{town} the $l$ parameter was interpreted as proportional to the
Planck length $l_{P}$, which provides an intuitive way to understand the
translation group as a non-abelian group on the Planck scale ($\sim 10^{-33}$%
cm) which appears as an abelian group on larger scales such as the scale of
elementary particles ($\sim 10^{-13}$cm). In this same reference it was
found that for that a gravitational theory, understood as a gauge theory of
the de-Sitter group, contain the Einstein-Hilbert-Cartan action, it is
necessary that%
\begin{equation}
l^{2}=16\pi G/c^{4}.  \label{siete}
\end{equation}
This means that if the structure of space-time at the microscopic scale is
governed by the de-Sitter group, then the constant $l$ appears naturally as
a constant associated with gravitational interaction. From (\ref{siete}) and
from the definition of the Planck length it is straightforward to see that $%
l $ is proportional to the Planck length.

Consistency with the dual procedure of $S$-expansion in terms of the
Maurer-Cartan forms \cite{salg3} demands that $h^{A}$ inherits the unit of
length from the f\"{u}nfbein. That is why it is necessary to introduce the $%
l $ parameter again, this time associated with $h^{A}.$ Could be interesting
to observe that $\boldsymbol{J}_{AB}$ are still Lorentz generators, but $%
\boldsymbol{P}_{A}$ are no longer $AdS$ boosts. In fact, $\left[ P_{A},P_{B}%
\right] =Z_{AB}$. However $e^{A}$ still transform as a vector under Lorentz
transformations, as it must, in order to recover gravity in this scheme.

A Chern-Simons lagrangian in $d=5$ dimensions is defined to be the following
local function of a one-form gauge connection $\boldsymbol{A}$: 
\begin{equation}
L_{ChS}^{\left( 5\right) }\left( \boldsymbol{A}\right) =C\left\langle 
\boldsymbol{AF}^{2}-\frac{1}{2}\boldsymbol{A}^{3}\boldsymbol{F+}\frac{1}{10}%
\boldsymbol{A}^{5}\right\rangle ,  \label{lcs}
\end{equation}%
where $\left\langle \cdots \right\rangle $ denotes a invariant tensor for
the corresponding Lie algebra$,$ $F=dA+AA$ is the corresponding the two-form
curvature and $C$ is a constant \cite{zan}.

Using theorem~VII.2 of Ref.~\cite{salg2}, it is possible to show that the
only non-vanishing components of an invariant tensor for the $\mathfrak{B}%
_{5}$ algebra are given by%
\begin{eqnarray*}
\left\langle \boldsymbol{J}_{A_{1}A_{2}}\boldsymbol{J}_{A_{3}A_{4}}\mathbf{P}%
_{A_{5}}\right\rangle &=&\alpha _{1}\frac{4l^{3}}{3}\varepsilon _{A_{1}\cdot
\cdot \cdot A_{5}}, \\
\left\langle \boldsymbol{J}_{A_{1}A_{2}}\boldsymbol{J}_{A_{3}A_{4}}\mathbf{Z}%
_{A_{5}}\right\rangle &=&\alpha _{3}\frac{4l^{3}}{3}\varepsilon _{A_{1}\cdot
\cdot \cdot A_{5}}, \\
\left\langle \boldsymbol{J}_{A_{1}A_{2}}\boldsymbol{Z}_{A_{3}A_{4}}\mathbf{P}%
_{A_{5}}\right\rangle &=&\alpha _{3}\frac{4l^{3}}{3}\varepsilon _{A_{1}\cdot
\cdot \cdot A_{5}},
\end{eqnarray*}%
where $\alpha _{1}$ and $\alpha _{3}$ are arbitrary independent constants of
dimensions $\left[ length\right] ^{-3}.$

Using the extended Cartan's homotopy formula as in Ref. \cite{salg4}, and
integrating by parts, it is possible to write down the Chern-Simons
Lagrangian in five dimensions for the $\mathfrak{B}_{5}$ algebra as \cite%
{salg1}, \cite{cuatro}

\begin{align}
L_{\text{EChS}}^{(5)}& =\alpha _{1}l^{2}\varepsilon _{ABCDE}e^{A}R^{BC}R^{DE}
\notag \\
& \quad +\alpha _{3}\varepsilon _{ABCDE}\left( \frac{2}{3}%
R^{AB}e^{C}e^{D}e^{E}+l^{2}R^{AB}R^{CD}h^{E}+2l^{2}k^{AB}R^{CD}\mathbf{T}%
^{E}\right)  \notag \\
& +d\hat{B}_{EChS}^{(4)},  \label{3}
\end{align}%
with 
\begin{eqnarray}
\hat{B}_{EChS}^{(4)} &=&\varepsilon _{abcde}\left\{ \alpha
_{1}l^{2}e^{a}\omega ^{bc}\left( \frac{2}{3}\omega ^{de}+\frac{1}{2}\omega
^{d}\,_{f}\omega ^{fe}\right) +\right.  \notag \\
&&+\alpha _{3}l^{2}\left( \left( h^{a}\omega ^{bc}+k^{ab}e^{c}\right) \left( 
\frac{2}{3}\omega ^{de}+\frac{1}{2}\omega ^{d}\,_{f}\omega ^{fe}\right)
+\right.  \notag \\
&&\left. +\left( k^{ab}\omega ^{cd}\left( \frac{2}{3}de^{e}+\frac{1}{2}%
\omega ^{e}\,_{f}e^{f}\right) \right) +\frac{\alpha _{3}}{6}%
e^{a}e^{b}e^{c}\omega ^{de}\right\}  \label{3'}
\end{eqnarray}%
where $\alpha _{1}$, $\alpha _{3}$ are parameters of the theory, $l$ is a
coupling constant, $R^{AB}=\text{d}\omega ^{AB}+\omega _{\text{ \ }%
C}^{A}\omega ^{CB}$ and $T^{A}=De^{A}$ correspond to the curvature $2$-form
and the torsion $2$-form respectively in the first-order formalism related
to the spin connection $1$-form and $e^{A}$, $h^{A}$ and $k^{AB}$ are others
gauge fields presents in the theory. It should be noted that the kinetic
terms for the $h^{A}$ and $k^{AB}$ fields are present only in the surface
term of the Lagrangian shown in (\ref{3'}).

It is also interesting to note that when the constant $\alpha _{1}$
vanishes, the lagrangian (\ref{3}) almost exactly matches the one given in
Ref.~\cite{edel}, the only difference being that in our case the coupling
constant $l^{2}$ appears explicitly in the last two terms. The presence or
absence of the coupling constant $l$ in the lagrangian could seem like a
minor or trivial matter, but it is not. As the authors of Ref.~\cite{edel}
clearly state, the presence of the Einstein--Hilbert term in this kind of
action does not guarantee that the dynamics will be that of general
relativity. In general, extra constraints on the geometry do appear, even
around a "vacuum" solution with $k^{ab}=h^{a}=0$. In fact, the variation of
the lagrangian, modulo boundary terms, can be written as%
\begin{eqnarray}
\delta L_{\mathrm{CS}}^{\left( 5\right) } &=&\varepsilon _{abcde}\left(
2\alpha _{3}R^{ab}e^{c}e^{d}+\alpha _{1}l^{2}R^{ab}R^{cd}+2\alpha _{3}l^{2}%
\mathrm{D}_{\omega }k^{ab}R^{cd}\right) \delta e^{e}  \notag \\
&&+\alpha _{3}l^{2}\epsilon _{abcde}R^{ab}R^{cd}\delta h^{e}+  \notag \\
&&+2\epsilon _{abcde}\delta \omega ^{ab}\left( \alpha
_{1}l^{2}R^{cd}T^{e}+\alpha _{3}l^{2}\mathrm{D}k^{cd}T^{e}+\alpha
_{3}e^{c}e^{d}T^{e}\right.  \notag \\
&&\left. +\alpha _{3}l^{2}R^{cd}\mathrm{D}h^{e}+\alpha _{3}l^{2}R^{cd}k_{%
\text{ \ }f}^{e}e^{f}+2\alpha _{3}l^{2}\epsilon _{abcde}\delta
k^{ab}R^{cd}T^{e}\right) .
\end{eqnarray}

This means that when the condition $\alpha _{1}=0$ is chosen, the
torsionless condition imposed, and a solution without matter $%
(k^{ab}=h^{a}=0 $) is picked out, we are left with%
\begin{equation}
\delta L_{\mathrm{CS}}^{\left( 5\right) }=2\alpha _{3}\epsilon
_{abcde}R^{ab}e^{c}e^{d}\delta e^{e}+\alpha _{3}l^{2}\varepsilon
_{abcde}R^{ab}R^{cd}\delta h^{e}.
\end{equation}

In this way, besides general relativity equations of motions $\epsilon
_{abcde}R^{ab}e^{c}e^{d}=0,$ the equations of motion of pure Gauss-Bonnet
theory $\varepsilon _{abcde}R^{ab}R^{cd}=0$ do also appear as an anomalous
constraint on the geometry. It is at this point where the presence of the
coupling constant $l$ makes the difference. In the present approach, it does
play the role of a coupling constant between geometry and \textquotedblleft
matter\textquotedblright . For this reason, in this case the limit $%
l\rightarrow 0$ leads to the Einstein--Hilbert term in the lagrangian, 
\begin{equation}
L_{\mathrm{CS}}^{\left( 5\right) }=\frac{2}{3}\alpha _{3}\varepsilon
_{abcde}R^{ab}e^{c}e^{d}e^{e}.
\end{equation}

In the same way, when we impose the weak limit of coupling constant, $%
l\rightarrow 0,$ the extra constraints just vanish, and $\delta L_{\mathrm{CS%
}}^{\left( 5\right) }=0$ lead us to just the Einstein--Hilbert dynamics in
the vacuum,%
\begin{equation}
\delta L_{\mathrm{CS}}^{\left( 5\right) }=2\alpha _{3}\varepsilon
_{abcde}R^{ab}e^{c}e^{d}\delta e^{e}+2\alpha _{3}\varepsilon _{abcde}\delta
\omega ^{ab}e^{c}e^{d}T^{e}.
\end{equation}

However, the Einstein-Chern-Simons gravity is valid only in odd dimensions
and in order to have a well defined four-dimensional theory is necessary to
carry out a dimensional reduction.

The aim of the present work is to derive the efective Einstein-Chern-Simons
equations on the 3-brane following the procedure used in Ref. \cite{br8}.
For simplicity the bulk spacetime is assumed to have 5 dimensions. In the
beginning we do not assume any conditions on the bulk spacetime. Later, we
assume the $Z_{2}$ symmetry and confinement of the matter energy momentum
tensor on the brane, in accordance with the brane world scenario.

This work is organized as follows: in section $2$ we briefly review the
Izrael's junction condition for Lovelock and AdS-Chern-Simons gravity. In
section $3$ we will study the junction conditions of Israel for the case of
the Einstein-Chern-Simons equations. In section $4$ we will obtain $4$%
-dimensional effective gravitational equations on the brane. Conclusions and
discussion are presented in Sect. $5$. Appendices present details omitted in
the main text.

\section{\textbf{Israel's junction condition for Lovelock and
AdS-Chern-Simons gravity}}

To study the braneworld in the context of Einstein-Chern-Simons gravity it
is necessary first know the junction conditions in the context of the
Lovelock gravity theory \cite{Lovelock}, which allow us to find the
appropriate junction conditions for AdS-Chern-Simons gravity.

In Refs. \cite{York},\cite{hawk},\cite{myers}, the study of the general
theory of relativity was generalized to the case of an edged manifold. This
has consequences for the application of the variational principle. In fact,
the total derivative term in the Euler-Lagrange variation leads to an
boundary term over $\partial M$.

The variation of an action of the form $S=S\left( g_{AB},\partial
_{C}g_{AB},\partial _{C}\partial _{D}g_{AB}\right) $ leads to~\cite{will} 
\begin{eqnarray}
\delta S &=&\int \mathrm{d}x^{D}\left\{ \left[ \frac{\partial \mathcal{L}}{%
\delta g^{AB}}-\frac{\partial \mathcal{L}}{\partial \left( \partial
_{C}g^{AB}\right) }+\partial _{D}\partial _{C}\left( \frac{\partial \mathcal{%
L}}{\partial \left( \partial _{D}\partial _{C}g^{AB}\right) }\right) \right]
\delta g^{AB}\right.  \notag \\
&&+\left. \partial _{C}\left[ \frac{\partial \mathcal{L}}{\partial \left(
\partial _{C}g^{AB}\right) }\delta g^{AB}-\partial _{D}\left( \frac{\partial 
\mathcal{L}}{\partial \left( \partial _{D}\partial _{C}g^{AB}\right) }%
\right) \delta g^{AB}+\partial _{D}\delta g^{AB}\frac{\partial \mathcal{L}}{%
\partial \left( \partial _{C}\partial _{D}g^{AB}\right) }\right] \right\} . 
\notag \\
&&  \label{5}
\end{eqnarray}

The last term corresponds to a boundary term. If there were only terms
proportional to $\delta g^{\mu \nu }$ and proportional to the derivative of $%
\delta g^{\mu \nu }$ at the boundary, there would be no problem. The problem
arises when there are normal derivatives of $\delta g^{\mu \nu }$ at the
boundary coming from the last term of~(\ref{5}). In other words, the
problematic terms are the ones proportional to $\partial _{D}\delta g^{AB}$
at the boundary because they cannot be overridden by fixing the induced
metric on the hypersurface. For this reason, it is necessary to add a term
to the Lagrangian that cancels the contribution of the term proportional to $%
\partial _{D}\delta g^{AB}$ at the edge, i.e., it cancels the normal
derivatives of the metric variation. In Lovelock gravity, R. C. Myers
proposed an appropriate boundary term in Ref.~\cite{myers}, and Willinson
and Gravanis generalized it in Refs.~\cite{will,gravanis1}. The term they
proposed corresponds to%
\begin{equation}
S_{\Sigma }=\sum_{n}n\beta _{n}\int_{\Sigma }\int_{0}^{1}\mathrm{d}%
t\,\varepsilon _{A_{1}.A_{2n}A_{2n+1}..A_{d}}\Theta
^{A_{1}A_{2}}R_{(t)}^{A_{3}A_{4}}..R_{(t)}^{A_{2n-1}A_{2n}}\bar{e}%
^{A_{2n+1}}...\bar{e}^{A_{d}},  \label{gravanis}
\end{equation}%
where $\beta _{n}$ are arbitrary constants, $R_{(t)}^{AB}=\mathrm{d}\omega
_{(t)}^{AB}+\omega ^{A}\,_{(t)L}\omega _{(t)}^{LB}$ is the curvature $2$%
-form, $\omega _{(t)}^{AB}=\bar{\omega}^{AB}+t\Theta ^{AB}$ is the $1$-form
that interpolates between the bulk spin connection and the one of the brane;
here $\Theta ^{AB}=\omega ^{AB}-\bar{\omega}^{AB}$.

\subsection{\textbf{Boundary term\ for AdS-Chern-Simons Lagrangian}}

In references \cite{mora1,mora2,mora3,mora4,oliv,kof}, was constructed a
boundary term that regularizes the action for AdS Chern-Simons gravity in ($%
2n+1$)- dimensions. In Chern-Simons AdS gravity, the Lagrangian is
constructed from Euler's topological invariant for the group $SO(2n,2)$, so
that the action is given by 
\begin{equation}
S_{\mathrm{2n+1}}^{\mathrm{A}\left( \mathrm{dS}\right) }=\int_{M_{2n+1}}L_{%
\mathrm{2n+1}}^{\mathrm{A}\left( \mathrm{dS}\right) }+\int_{\Sigma =\partial
M_{2n+1}}B_{2n}  \label{cs1}
\end{equation}%
where%
\begin{eqnarray}
L_{\mathrm{2n+1}}^{\mathrm{A}\left( \mathrm{dS}\right) }
&=&\int_{0}^{1}dt\left\langle F_{t}^{n}e\right\rangle
=\int_{0}^{1}dt\varepsilon _{A_{1}\cdots
A_{2n+1}}F_{t}^{A_{1}A_{2}}F_{t}^{A_{3}A_{4}}\cdots
F_{t}^{A_{2n-1}A_{2n}}e^{A_{2n+1}}  \notag \\
B_{2n} &=&-n(n+1)\int_{0}^{1}ds\int_{0}^{1}dt\text{ }s\text{ }\left\langle
A_{t}\Theta \left( sF_{t}+s(s-1)A_{t}^{2}\right) ^{n-1}\right\rangle
\label{cs2}
\end{eqnarray}%
with 
\begin{equation}
A_{t}=tA+(1-t)\bar{A}=\bar{A}+t\Theta ,\text{ with }\Theta =A-\bar{A}
\label{chs1}
\end{equation}%
\begin{eqnarray}
F_{t} &=&dA_{t}+A_{t}^{2}=tF+(1-t)\bar{F}-t(1-t)\Theta ^{2}  \notag \\
&=&dA+\left( t-1\right) d\Theta +\frac{1}{2}\left[ A,A\right] +\left(
t-1\right) \left[ A,\Theta \right] +\frac{\left( t-1\right) ^{2}}{2}\left[
\Theta ,\Theta \right] ,  \label{chs2}
\end{eqnarray}%
where $A$ corresponds to the $1$-form gauge potential of the hypersurface $%
\Sigma $ and $A$ is the $1$-form gauge potential of the bulk.

Let us now consider the $5$-dimensional case. In this case we have

\begin{equation}
S_{\mathrm{5}}^{\mathrm{A}\left( \mathrm{dS}\right) }=\int_{M_{2n+1}}L_{%
\mathrm{5}}^{\mathrm{A}\left( \mathrm{dS}\right) }+\int_{\partial
M_{2n+1}}B_{4},  \label{chs3}
\end{equation}%
where $L_{\mathrm{5}}^{\mathrm{A}\left( \mathrm{dS}\right) }$ is given by
the first equation in (\ref{cs2}) when $n=2$ and $B_{4}$ is given by 
\begin{eqnarray}
B_{4} &=&-6\int_{0}^{1}dt\int_{0}^{1}ds\text{ }\left\langle A_{t}\Theta
\left( s^{2}F_{t}+s^{2}(s-1)A_{t}^{2}\right) \right\rangle  \notag \\
&=&\int_{0}^{1}dt\text{ }\left\langle \frac{1}{2}\Theta A_{t}^{3}-2\Theta
A_{t}F_{t}\right\rangle .  \label{cs3}
\end{eqnarray}%
introducing (\ref{chs1},\ref{chs2}) in (\ref{cs3}) we find

\begin{eqnarray}
B_{4} &=&\left\langle \Theta A\left( -\frac{3}{2}F+d\Theta -\frac{1}{2}dA+%
\frac{3}{4}\left[ A,\Theta \right] -\frac{1}{4}\left[ \Theta ,\Theta \right]
\right) +\Theta \Theta \left( \frac{3}{4}F-\frac{2}{3}d\Theta +\right.
\right.  \notag \\
&&\left. \left. +\frac{1}{4}dA-\frac{1}{2}\left[ A,\Theta \right] +\frac{3}{%
16}\left[ \Theta ,\Theta \right] \right) \right\rangle ,  \label{cs4}
\end{eqnarray}%
where the gauge potential $1$-form as well as their corresponding curvature $%
2$-form for the AdS algebra, are given by 
\begin{eqnarray}
A &=&\frac{1}{l}e^{A}\tilde{P}_{A}+\frac{1}{2}\omega ^{AB}\tilde{J}_{AB} 
\notag \\
F &=&\frac{1}{l}\mathbf{T}^{A}\tilde{P}_{A}+\frac{1}{2}\left( R^{AB}+\frac{1%
}{l^{2}}e^{A}e^{B}\right) \tilde{J}_{AB}  \notag \\
\bar{A} &=&\frac{1}{l}\bar{e}^{A}\tilde{P}_{A}+\frac{1}{2}\bar{\omega}^{AB}%
\tilde{J}_{AB},  \notag \\
\bar{F} &=&\frac{1}{l}\bar{\mathbf{T}}^{A}\tilde{P}_{A}+\frac{1}{2}\left( 
\bar{R}^{AB}+\frac{1}{l^{2}}\bar{e}^{A}\bar{e}^{B}\right) \tilde{J}_{AB}.
\label{cs5}
\end{eqnarray}%
Here $e^{A}$ and $\bar{e}^{A}$ are the vierbeins, $\omega ^{AB}$ and $\bar{%
\omega}^{AB}$ are the spin connections, $R^{AB}=d\omega ^{AB}+\omega _{\text{
\ }C}^{A}\omega ^{CB}$ and $\bar{R}^{AB}=d\bar{\omega}^{AB}+\bar{\omega}_{%
\text{ }C}^{A}\bar{\omega}^{CB}$ are the curvature $2$-forms. $\tilde{P}_{A} 
$ and $\tilde{J}_{AB}$ are the generators of the AdS algebra.

Taking into account that considering the non-zero components of the
invariant tensor for the AdS algebra in $5$ dimensions are proportional to
the Levi-Civita symbol, namely, 
\begin{equation}
\left\langle \tilde{J}_{AB}\tilde{J}_{CD}\tilde{P}_{E}\right\rangle =\frac{%
4\kappa }{3}\varepsilon _{ABCDE}  \label{cs7}
\end{equation}%
we have that terms of order greater than or equal to $2$ in $\Theta $ will
be null, so that (\ref{cs3}) takes the form

\begin{equation}
B_{4}=\left\langle \Theta A\left( -\frac{3}{2}F-\frac{1}{2}dA+\frac{3}{4}%
\left[ A,\Theta \right] -\frac{1}{4}\left[ \Theta ,\Theta \right] \right)
\right\rangle ,  \label{cs8}
\end{equation}%
where, using the commutation relations of the AdS algebra, we find 
\begin{eqnarray}
\left[ \Theta ,A\right] &=&\omega ^{A}\,_{L}\Theta ^{LB}J_{AB}+\frac{2}{l}%
\Theta ^{A}\,_{L}e^{L}P_{A}  \notag \\
\left[ \Theta ,\Theta \right] &=&\Theta ^{A}\,_{L}\Theta ^{LB}J_{AB}.
\label{cs9}
\end{eqnarray}%
Introducing (\ref{cs9}) in (\ref{cs8}) and using (\ref{cs7})\ \ we obtain

\begin{eqnarray}
B_{4}^{AdS} &=&\frac{\kappa }{3l}\varepsilon _{ABCDE}\left\{ -\frac{1}{2}%
\Theta ^{AB}\omega ^{CD}\mathbf{T}^{E}+\frac{3}{2}D\Theta ^{AB}\omega
^{CD}e^{E}-\frac{7}{2}\Theta ^{AB}R^{CD}e^{E}+\right.  \notag \\
&&\left. -\omega ^{A}\,_{L}\Theta ^{LB}\omega ^{CD}e^{E}+\frac{3}{2}\Theta
^{AB}\omega ^{C}\,_{L}\omega ^{LD}e^{E}-\frac{1}{2}\Theta ^{AB}\Theta
^{C}\,_{L}\Theta ^{LD}e^{E}\right\}  \label{cs11}
\end{eqnarray}%
where $T^{E}$ is the torsion $2$-form.

In the braneworld context the $A$ indices run from $0$ to $4$ and the $a$
indices run from $0$ to $3$. This means that $\omega ^{AB}=\left( \omega
^{ab},\omega ^{a4}\right) ,$ $e^{A}=\left( e^{a},e^{4}\right) ,$ $\omega
^{ab}=\bar{\omega}^{ab}$ and $e^{a}=\bar{e}^{a}$. So keeping in mind that $%
\Theta ^{AB}=\omega ^{AB}-\bar{\omega}^{AB}$ and $\Theta ^{A}=e^{A}-\bar{e}%
^{A}$ we can write 
\begin{equation}
\Theta =\frac{1}{2}\Theta ^{AB}\tilde{J}_{AB}+\frac{1}{l}\Theta ^{A}\tilde{P}%
_{A}=\omega ^{4a}\tilde{J}_{4a}+\frac{1}{l}e^{4}\tilde{P}_{4}.  \label{cs10}
\end{equation}

Introducing these results in (\ref{cs11}) we find 
\begin{eqnarray}
B_{4}^{AdS} &=&\kappa \varepsilon _{abcd}\left\{ -\frac{1}{3l}\omega
^{4a}\omega ^{bc}\mathbf{T}^{d}-\frac{7}{3l}\omega ^{4a}\left( R^{bc}+\frac{1%
}{l^{2}}e^{b}e^{c}\right) e^{d}+\frac{4}{3l^{3}}\omega
^{4a}e^{b}e^{c}e^{d}+\right.  \notag \\
&&\left. +\frac{1}{l}R^{4a}\omega ^{bc}e^{d}-\frac{1}{3}\omega
^{a}\,_{l}\omega ^{l4}\omega ^{bc}e^{d}\right\}  \label{cs12}
\end{eqnarray}

\section{\textbf{Israel junction condition for} \textbf{%
Einstein-Chern-Simons gravity}}

\subsection{\textbf{Boundary term\ for Einstein-Chern-Simons Lagrangian}}

In the introduction we noted that the so-called Einstein-Chern-Simons
Lagrangian (\ref{3}) was obtained from the AdS-Chern-Simons Lagrangian by
means of the expansion procedure. The corresponding boundary term for the
EChS Lagrangian can be obtained from the boundary term of the
AdS-Chern-Simons Lagrangian following the same procedure. Indeed, making use
of the dual S-expansion procedure \cite{salg3}, it is found that the
boundary term for the Einstein-Chern-Simons Lagrangian (\ref{3}), and that
must be added to it, is given by 
\begin{eqnarray}
B_{4}^{EChS} &=&\varepsilon _{abcd}\left\{ \alpha _{1}l^{2}\left( -\frac{1}{3%
}\omega ^{4a}\omega ^{bc}\mathbf{T}^{d}-\frac{7}{3}\omega
^{4a}R^{bc}e^{d}+R^{4a}\omega ^{bc}e^{d}-\frac{1}{3}\omega ^{a}\,_{l}\omega
^{l4}\omega ^{bc}e^{d}\right) +\right.  \notag \\
&&+\alpha _{3}l^{2}\left( -\frac{1}{3}\omega ^{4a}\omega ^{bc}Dh^{d}-\frac{7%
}{3}\omega ^{4a}R^{bc}h^{d}+R^{4a}\omega ^{bc}h^{d}-\frac{1}{3}\omega _{%
\text{ \ }l}^{a}\omega ^{l4}\omega ^{bc}h^{d}\right) +  \notag \\
&&\left. -\alpha _{3}\omega ^{4a}e^{b}e^{c}e^{d}\right\} ,  \label{cs13}
\end{eqnarray}%
where we have used $k^{AB}=0$ and $h^{4}=0$.

This means that the action for so-called Einstein-Chern-Simons gravity in
five dimensions is given by

\begin{eqnarray}
S_{EChS}^{(5)} &=&\int_{M}\varepsilon _{ABCDE}\left[ \alpha
_{1}l^{2}R^{AB}R^{CD}e^{E}+\alpha _{3}\left( \frac{2}{3}%
R^{AB}e^{C}e^{D}e^{E}+l^{2}R^{AB}R^{CD}h^{E}\right) \right]  \notag \\
&&+\int_{\Sigma }\varepsilon _{abcd}\left\{ \alpha _{1}l^{2}\left(
4K^{a}R^{bc}e^{d}+\frac{1}{3}K^{a}K^{b}K^{c}e^{d}\right) +\right.  \notag \\
&&+\alpha _{3}l^{2}\left( 4K^{a}R^{bc}h^{d}+\frac{1}{3}K^{a}K^{b}K^{c}h^{d}%
\right) +\left. \frac{4}{3}\alpha _{3}K^{a}e^{b}e^{c}e^{d}\right\} ,
\label{cs14}
\end{eqnarray}%
where we have used the fact that when the torsion is zero, it is possible to
write the normal component of the spin connection in the form\textbf{\ }%
\begin{equation}
\omega ^{4a}=-K^{a}=-K^{a}\,_{l}e^{l},  \label{cs15}
\end{equation}%
where $K^{a}\,_{l}$\ corresponds to the extrinsic curvature.

Taking the limit $l\rightarrow 0$ (low energy limit) we find

\begin{equation}
S_{EChS}^{(5)}=\frac{2}{3}\alpha _{3}\int_{M}\varepsilon
_{ABCDE}R^{AB}e^{C}e^{D}e^{E}+\frac{4}{3}\alpha _{3}\int_{\Sigma
}\varepsilon _{abcd}K^{a}e^{b}e^{c}e^{d},  \label{cs16}
\end{equation}%
which corresponds to the Einstein-Hilbert term plus the
Gibbons-Hawkings-York boundary term, that is, in the limit $l\rightarrow 0$
general relativity is recovered.

\subsection{\textbf{Einstein-Chern-Simons junction}}

Since the brane divides the space $M$ into two spaces, $M^{+}$ with metric $%
g_{\mu \nu }^{+}$ and $M^{-}$ with metric $g_{\mu \nu }^{-}$. Each of these
spaces induces a metric $q_{ab}^{\pm }$ on the brane, which has associated a
normal vector pointing from $M^{-}$ to $M^{+}$. Due to this division the
total action can be written in the form

\begin{eqnarray}
S_{total}^{EChS(5)} &=&S_{EChS+}^{(5)}-S_{EChS-}^{(5)}+\int_{M}L_{M}  \notag
\\
&=&\int_{M^{+}}\varepsilon _{ABCDE}\left( \alpha
_{1}l^{2}R_{+}^{AB}R_{+}^{CD}e_{+}^{E}+\alpha _{3}\left( \frac{2}{3}%
R_{+}^{AB}e_{+}^{C}e_{+}^{D}e_{+}^{E}+l^{2}R_{+}^{AB}R_{+}^{CD}h_{+}^{E}%
\right) \right)  \notag \\
&&+\int_{M^{-}}\varepsilon _{ABCDE}\left( \alpha
_{1}l^{2}R_{-}^{AB}R_{-}^{CD}e_{-}^{E}+\alpha _{3}\left( \frac{2}{3}%
R_{-}^{AB}e_{-}^{C}e_{-}^{D}e_{-}^{E}+l^{2}R_{-}^{AB}R_{-}^{CD}h_{-}^{E}%
\right) \right)  \notag \\
&&+\int_{\Sigma }\varepsilon _{abcd}\left\{ \alpha _{1}l^{2}\left( 4\left[
K^{a}\right] R^{bc}e^{d}+\frac{1}{3}\left[ K^{a}K^{b}K^{c}\right]
e^{d}\right) +\right.  \notag \\
&&\left. +\alpha _{3}l^{2}\left( 4\left[ K^{a}\right] R^{bc}h^{d}+\frac{1}{3}%
\left[ K^{a}K^{b}K^{c}\right] h^{d}\right) +\frac{4}{3}\alpha _{3}\left[
K^{a}\right] e^{b}e^{c}e^{d}\right\}  \notag \\
&&+\kappa \int_{M}L_{M}  \label{act1}
\end{eqnarray}%
where $S_{EChS\pm }$ corresponds to the action of the bulk in each part of
the space, $L_{M},$ correspond to matter contribution to the action and $%
\left[ X\right] =X_{+}-X_{-}$. In the previous action we have considered
that the metric of the space is continuous on $\Sigma $ in such a way that
the connection is well defined from the metric. We have also considered that
the field $h^{A}$ is different from zero only in the brane, that is

\begin{equation}
h^{4}=0;\text{ \ \ }h^{a}=\bar{h}^{a}.
\end{equation}

Varying (\ref{act1}) we have 
\begin{eqnarray}
\delta S_{total}^{EChS(5)} &=&\int_{M^{+}}\varepsilon _{ABCDE}\left\{ \left(
\alpha _{1}l^{2}R_{+}^{AB}R_{+}^{CD}+2\alpha
_{3}R_{+}^{AB}e_{+}^{C}e_{+}^{D}\right) \delta e_{+}^{E}+\right.  \notag \\
&&\left. +\alpha _{3}l^{2}R_{+}^{AB}R_{+}^{CD}\delta h_{+}^{E}+2\alpha
_{3}R_{+}^{AB}Dh^{C}\delta \omega _{+}^{DE}\right\} +  \notag \\
&&+\int_{M^{-}}\varepsilon _{ABCDE}\left\{ \left( \alpha
_{1}l^{2}R_{-}^{AB}R_{-}^{CD}+2\alpha
_{3}R_{-}^{AB}e_{-}^{C}e_{-}^{D}\right) \delta e_{-}^{E}+\right.  \notag \\
&&\left. +\alpha _{3}l^{2}R_{-}^{AB}R_{-}^{CD}\delta h_{-}^{E}+2\alpha
_{3}R_{-}^{AB}Dh_{-}^{C}\delta \omega _{-}^{DE}\right\} +  \notag \\
&&+\int_{\Sigma }\varepsilon _{abcd}\left\{ \left( 4\alpha _{1}l^{2}\left[
K^{a}\right] \bar{R}^{bc}+\frac{13}{3}\alpha _{1}l^{2}\left[ K^{a}K^{b}K^{c}%
\right] \right. \right. +  \notag \\
&&\left. +\frac{4}{3}\alpha _{3}\left[ K^{a}\right] e^{b}e^{c}\right) \delta
e^{d}+\alpha _{3}l^{2}\left( 4\left[ K^{a}\right] \bar{R}^{bc}+\frac{13}{3}%
\left[ K^{a}K^{b}K^{c}\right] \right) \delta h^{d}+  \notag \\
&&+\left( 9l^{2}K_{+}^{a}K_{+}^{b}\left( \alpha _{1}e^{c}+\alpha
_{3}h^{c}\right) +12\alpha _{3}e^{a}e^{b}e^{c}\right) \delta K_{+}^{d}+ 
\notag \\
&&+\left( 9l^{2}K_{-}^{a}K_{-}^{b}\left( \alpha _{1}e^{c}+\alpha
_{3}h^{c}\right) +12\alpha _{3}e^{a}e^{b}e^{c}\right) \delta K_{-}^{d}+ 
\notag \\
&&+\kappa \int_{M}\left( T_{E}\delta e^{E}+T_{E}^{(h)}\delta h^{E}\right) =0,
\label{act2}
\end{eqnarray}%
where $T_{E}=-\frac{1}{4!}\varepsilon
_{ABCDS}T^{S}\,_{E}e^{A}e^{B}e^{C}e^{D} $ and $T_{E}^{(h)}=-\frac{1}{4!}%
\varepsilon _{ABCDS}T^{(h)S}\,_{E}e^{A}e^{B}e^{C}e^{D}$ corresponds to the
forms energy-momentum associated to $e^{A}$ y $h^{A}$, respectively.

The previous structure allows to separate the energy-momentum tensors in the
form

\begin{equation}
T_{E}=Q_{+E}\Theta \left( \chi \right) +Q_{-E}\Theta \left( -\chi \right)
+\delta \left( \chi \right) \bar{T}_{E},
\end{equation}%
where $\chi $ is a coordinate defined in the direction normal to $\Sigma $
such that the brane is located at $\chi =0$. So $\bar{T}_{d}=-\frac{1}{3!}%
\varepsilon _{abcs}\bar{T}^{s}\,_{d}e^{a}e^{b}e^{c}$, is the part of
energy-momentum tensor on the brane and $Q_{E}$ the part external to the
brane (bulk). Bearing this in mind, the equation (\ref{act2}) leads to the
following field equations 
\begin{eqnarray}  \label{bran2}
4\kappa _{5}Q_{\pm E} &=&\varepsilon _{ABCDE}\left( \alpha _{1}l^{2}R_{\pm
}^{AB}R_{\pm }^{CD}+2\alpha _{3}R_{\pm }^{AB}e_{\pm }^{C}e_{\pm }^{D}\right)
,  \label{em1} \\
8\kappa _{5}Q_{\pm E}^{(h)} &&\alpha _{3}l^{2}\varepsilon _{ABCDE}R_{\pm
}^{AB}R_{\pm }^{CD},  \label{em2} \\
0 &=&2\alpha _{3}\varepsilon _{ABCDE}R_{\pm }^{AB}Dh_{\pm }^{C},  \label{em3}
\\
\kappa \bar{T}_{d} &=&\varepsilon _{abcd}\left( 4\alpha _{1}l^{2}\left[ K^{a}%
\right] \bar{R}^{bc}+\frac{13}{3}\alpha _{1}l^{2}\left[ K^{a}K^{b}K^{c}%
\right] \right. +  \notag \\
&&\left. +\frac{4}{3}\alpha _{3}\left[ K^{a}\right] e^{b}e^{c}\right) ,
\label{em4} \\
\kappa \bar{T}_{d}^{(h)} &=&\alpha _{3}l^{2}\varepsilon _{abcd}\left( 4\left[
K^{a}\right] \bar{R}^{bc}+\frac{13}{3}\left[ K^{a}K^{b}K^{c}\right] \right) ,
\label{em5} \\
0 &=&\varepsilon _{abcd}\left( 9l^{2}K_{\pm }^{a}K_{\pm }^{b}\left( \alpha
_{1}e^{c}+\alpha _{3}h^{c}\right) +12\alpha _{3}e^{a}e^{b}e^{c}\right) ,
\label{em6}
\end{eqnarray}%
with $\kappa _{5}=\kappa /8\alpha _{3}$.

Note that the equations (\ref{em1}-\ref{em3}) correspond to the
five-dimensional EChS field equations and the equations (\ref{em4}-\ref{em6}%
) are the conditions that must be satisfied for that the curvature to be
well defined at the junction.

Replacing (\ref{em5}) in (\ref{em4}) we obtain

\begin{equation}
-\varepsilon _{abcd}\left[ K^{a}\right] e^{b}e^{c}=2\kappa _{5}\tilde{\bar{T}%
}_{d},  \label{cs22}
\end{equation}%
where $\tilde{\bar{T}}_{\text{ }d}^{k}=\bar{T}_{\text{ }d}^{k}\,+\alpha \bar{%
T}_{\text{ \ }d}^{k(h)}\,$.

It is possible to solve (\ref{cs22}) considering $K^{a}=K^{a}\,_{l}e^{l}$.
In fact, 
\begin{equation}
\varepsilon _{abcd}\left[ K^{a}\,_{l}\right] e^{l}e^{b}e^{c}=2\kappa
_{5}\varepsilon _{abcs}\tilde{\bar{T}}^{s}\,_{d}e^{a}e^{b}e^{c},
\label{cs23}
\end{equation}%
which in tensor language takes the form 
\begin{equation}
\left[ K\right] \delta _{d}^{m}-\left[ K^{m}\,_{d}\right] =\kappa _{5}\tilde{%
\bar{T}}^{m}\,_{d}.  \label{cs24}
\end{equation}%
Contracting the indices $m$ and $d$ we find that $\left[ K\right] =\kappa
_{5}\tilde{\bar{T}}/3$. So that

\begin{equation}
\left[ K^{m}\,_{d}\right] =-\kappa _{5}\left( \tilde{\bar{T}}^{m}\,_{d}-%
\frac{1}{3}\tilde{\bar{T}}\delta _{d}^{m}\right) ,  \label{cs25}
\end{equation}%
where we can see that when $l\rightarrow 0$, $\alpha \rightarrow 0$ we have 
\begin{equation}
\left[ K^{m}\,_{d}\right] =-\kappa _{5}\left( \bar{T}^{m}\,_{d}-\frac{1}{3}%
\bar{T}\delta ^{m}\,_{d}\right) ,  \label{cs26}
\end{equation}%
which coincides with the usual Israel's junction condition as long as $%
\kappa =\kappa _{5}$. This means that the Lanczos equation for
Einstein-Chern-Simons gravity (\ref{cs25}) is an Israel-type join condition,
even though the Lagrangian contains quadratic terms on the curvature. Unlike
the case of General Relativity, the extrinsic curvature is subject to the
condition (\ref{em5}) on both faces of the brane.

\section{\textbf{3-brane in Einstein-Chern-Simons gravity}}

It is possible to study the 3-brane world in Einstein-Chern-Simons gravity
using the results of the previous section and the procedure of Ref.~\cite%
{br8}. We must consider the induced metric $q_{ab}$ on the brane $\Sigma $
and a normal vector $n^{\mu }$ embedded in the five-dimensional bulk $M$
with metric $g_{\mu \nu }$.

The action~(\ref{3}) gives rise to the field equations \cite{gomez} 
\begin{eqnarray}
\varepsilon _{ABCDE}R^{AB}e^{C}e^{D} &=&4k_{5}\left( T_{E}+\alpha
T_{E}^{(h)}\right) ,  \label{ecs1} \\
\frac{l^{2}}{8k_{5}}\varepsilon _{ABCDE}R^{AB}R^{CD} &=&T_{E}^{(h)},
\label{ecs2} \\
\varepsilon _{ABCDE}R^{CD}Dh^{E} &=&0,  \label{ecs3}
\end{eqnarray}%
where the capital letters denote the bulk indices $\left\{ 0,1,2,3,4\right\} 
$, while the lowercase letters will denote brane indices $\left\{
0,1,2,3\right\} \mathrm{.}$ The matter Lagrangian gives rise to two
stress-energy tensors, $T_{E}=\delta L_{M}/\delta e^{E}$ and $%
T_{E}^{(h)}=\delta L_{M}/\delta h^{E}$.

To study the brane, we must consider the interior derivative concerning the
normal vector to the brane $i_{A}$. Since $i_{a}e^{b}=\delta _{b}^{a}$ and $%
i_{4}R^{ab}e^{4}=0$, eq.~(\ref{ecs1}) takes the form%
\begin{equation}
\varepsilon _{a4cde}\left( i_{4}R^{a4}e^{c}-R^{ac}i_{4}e^{4}\right)
e^{d}=2k_{5}i_{4}\left( T_{e}+\alpha T_{e}^{(h)}\right) .  \label{br1}
\end{equation}

Here $i_{4}R^{a4}=-\Tilde{E}_{\hspace{5pt}m}^{a}e^{m}$ and\ $i_{4}R^{ac}=-%
\Tilde{B}_{\hspace{8pt}m}^{ac}e^{m}$, where $\Tilde{E}_{\hspace{5pt}m}^{a}$
\ and $\Tilde{B}_{\hspace{8pt}m}^{ac}$ are the so-called electric and
magnetic parts of the Riemann tensor.

For the second member of (\ref{br1}),%
\begin{equation}
i_{4}T_{e}=-\frac{1}{3!}\varepsilon _{abcm}T^{m}\,_{d}e^{a}e^{b}e^{c},
\label{br2}
\end{equation}%
and similarly for $T_{e}^{(h)}$. Therefore, eq.~(\ref{br1}) takes the form

\begin{equation}
2k_{5}i_{4}\tilde{T}_{d}=\varepsilon _{abcd}\left( \Tilde{E}_{\hspace{5pt}%
m}^{a}e^{m}e^{b}+\bar{R}^{ab}-K^{a}\,_{l}K_{m}^{b}e^{l}e^{m}\right) ,
\label{br3}
\end{equation}%
where we have used the Gaussian equation.

Now, let us consider eq.~(\ref{ecs2}). We have that

\begin{equation}
8k_{5}i_{4}T_{e}^{(h)}=4l^{2}\varepsilon _{4bcde}\left(
R^{4b}\,_{4m}e^{m}R^{cd}+R^{4b}R^{cd}\,_{4m}e^{m}\right) ,  \label{br4}
\end{equation}%
and using the Gauss-Codazzi equations, 
\begin{equation}
-2k_{5}i_{4}T_{d}^{(h)}=\varepsilon _{abcd}\left\{ \tilde{E}%
^{a}\,_{m}e^{m}\left( \bar{R}^{bc}-K^{b}\,_{l}K^{c}\,_{m}e^{l}e^{m}\right)
+R^{4a}\tilde{B}^{bc}\,_{m}e^{m}\right\} .  \label{br5}
\end{equation}

In the same way, from (\ref{ecs3}) we see that 
\begin{equation}
0=l^{2}\varepsilon _{abcd}\left\{ \tilde{E}^{c}\,_{l}e^{l}\mathcal{D}%
h^{d}+R^{4c}\mathcal{D}_{4}h^{d}\right\} .  \label{br7}
\end{equation}

On the other hand, from the decomposition of the curvature tensor 
\begin{equation}
R^{AB}\,_{CD}=C^{AB}\,_{CD}+\frac{2}{3}\left( \delta _{\lbrack
C}^{A}R^{B}\,_{D]}-\delta _{\lbrack C}^{B}R^{A}\,_{D]}\right) -\frac{1}{6}%
\delta _{\lbrack C}^{A}\delta _{D]}^{B}R,  \label{br8}
\end{equation}%
we can find $\Tilde{E}^{a}\,_{m}$ and $\tilde{B}^{ab}\,_{m}$ as a function
of the electric part $E^{a}\,_{m}$ and the magnetic part \ $B^{ab}\,_{m}$ of
the Weyl tensor. We have that $\Tilde{E}^{a}\,_{m}=R^{4a}\,_{4m}$, $\tilde{B}%
^{ab}\,_{m}=R^{ab}\,_{m4}$, $\ E^{a}\,_{m}=C^{4a}\,_{4m}$, and $\
B^{ab}\,_{m}=C^{ab}\,_{m4}$, and therefore,

\begin{eqnarray}
\Tilde{E}^{a}\,_{m} &=&E^{a}\,_{m}+\frac{1}{3}\left(
R^{b}\,_{d}+R^{4}\,_{4}\delta _{d}^{b}\right) -\frac{1}{12}R\delta _{a}^{b},
\label{br9} \\
\tilde{B}^{ab}\,_{m} &=&B^{ab}\,_{m}+\frac{1}{3}\left( \delta
_{c}^{a}R^{b}\,_{4}+\delta _{c}^{b}R^{a}\,_{4}\right) .  \label{br10}
\end{eqnarray}

Since 
\begin{eqnarray}
R &=&-\frac{2}{3}k_{5}\left( T+\alpha T^{(h)}\right) ,  \label{br11} \\
R^{b}\,_{d} &=&k_{5}\left( T^{b}\,_{d}+\alpha T^{(h)b}\,_{d}-\frac{1}{3}%
\left[ \alpha T^{(h)}+T\right] \delta _{d}^{b}\right) ,  \label{br12}
\end{eqnarray}%
we have that the relations between the electrical parts of the Riemann
tensor $\tilde{E}^{a}\,_{l}$ and the Weyl tensor $E^{a}\,_{l}$ are

\begin{equation}
\tilde{E}^{a}\,_{l}=E^{a}\,_{l}+\frac{k_{5}}{3}\left[ \tilde{T}%
^{a}\,_{l}+\delta _{l}^{a}\left( \tilde{T}^{4}\,_{4}-\frac{\tilde{T}}{2}%
\right) \right] ,  \label{elec1}
\end{equation}%
moreover, the relations between the corresponding magnetic parts are 
\begin{equation}
\tilde{B}^{ab}\,_{l}=B^{ab}\,_{l}+\frac{2}{3}k_{5}\delta _{l}^{[a}\tilde{T}%
^{b]}\,_{4}.  \label{magne1}
\end{equation}

Replacing (\ref{elec1}) and (\ref{magne1}) in the equations of motion, we
have

\begin{eqnarray}
\varepsilon _{abcd}R^{ab}e^{c} &=&-2k_{5}i_{4}\tilde{T}_{d}-\varepsilon
_{abcd}\left\{ \left( E^{a}\,_{m}e^{m}+\frac{k_{5}}{3}\left[ \tilde{T}%
^{a}\,_{l}e^{l}+\left( \tilde{T}^{4}\,_{4}-\frac{\tilde{T}}{2}\right) e^{a}%
\right] \right) e^{b}\right.  \notag \\
&&\left. -K^{a}\,_{l}K_{m}^{b}e^{l}e^{m}\right\} e^{c}  \label{eom1} \\
2k_{5}i_{4}\tilde{T}_{d}^{(h)} &=&-l^{2}\varepsilon _{abcd}\left(
E^{a}\,_{m}e^{m}+\frac{k_{5}}{3}\left[ \tilde{T}^{a}\,_{l}e^{l}+\left( 
\tilde{T}^{4}\,_{4}-\frac{\tilde{T}}{2}\right) e^{a}\right] \right) \bar{R}%
^{bc}+  \notag \\
&&-l^{2}\varepsilon _{abcd}\left\{ R^{4a}\left( B^{bc}\,_{l}e^{l}-\frac{2}{3}%
k_{5}^{b}\tilde{T}\,_{4}e^{c}\right)
-E^{a}\,_{m}e^{m}K^{b}\,_{l}K^{c}\,_{m}e^{l}e^{m}\right.  \notag \\
&&\left. -\frac{k_{5}}{3}\left[ \tilde{T}^{a}\,_{l}e^{l}+\left( \tilde{T}%
^{4}\,_{4}-\frac{\tilde{T}}{2}\right) e^{a}\right] K^{b}\,_{l}K^{c}%
\,_{m}e^{l}e^{m}\right\}  \label{eom2} \\
0 &=&l^{2}\varepsilon _{abcd}\left\{ R^{4c}\mathcal{D}_{4}h^{d}+\left( E_{%
\hspace{5pt}m}^{c}e^{m}+\frac{k_{5}}{3}\left( \tilde{T}^{c}\,_{l}e^{l}+%
\left( \tilde{T}^{4}\,_{4}+\right. \right. \right. \right.  \notag \\
&&\left. \left. \left. \left. -\frac{\tilde{T}}{2}\right) e^{c}\right)
\right) \mathcal{\mathcal{D}}_{\omega }h^{d}\right\}  \label{eom3}
\end{eqnarray}

The brane equations of motion are subject to the conditions given by the
normal component of the equations of motion in bulk. In the equations~(\ref%
{ecs1}-\ref{ecs3}) we make the free index equal to \textquotedblleft $4$%
\textquotedblright\ and then we apply the interior derivative with respect
to the normal, 
\begin{eqnarray}
l^{2}\varepsilon _{abcd}B^{ab}\,_{l}\bar{R}^{cd}e^{l}
&=&8k_{5}i_{4}T^{(h)}\,_{4}-l^{2}\varepsilon _{abcd}\left( \frac{2}{3}k_{5}%
\tilde{T}^{b}\,_{4}e^{a}K^{c}\,_{s}K^{d}\,_{m}e^{s}e^{m}+\right.  \notag \\
&&\left. +\frac{2}{3}k_{5}^{b}\tilde{T}\,_{4}e^{a}\bar{R}^{cd}-B^{ab}%
\,_{l}K^{c}\,_{s}K^{d}\,_{m}e^{l}e^{s}e^{m}\right)  \label{br13} \\
l^{2}\varepsilon _{abcd}R^{bc}D_{4}h^{d} &=&-l^{2}\varepsilon _{abcd}\left(
K^{b}\,_{l}K^{c}\,_{s}e^{l}e^{s}D_{4}h^{d}+\right.  \notag \\
&&\left. +B^{bc}\,_{l}\mathcal{D}h^{d}e^{l}+\frac{2}{3}k_{5}^{c}\tilde{T}%
\,_{4}e^{b}Dh^{d}\right)  \label{br14}
\end{eqnarray}

We have obtained these effective equations on the brane, and the conditions
given by the normal part of the equations of motion, using only the Gaussian
embedding, and without imposing additional conditions.

Following Ref. \cite{br8} we define the coordinate $\chi $, such that the
brane is at $\chi =0$. Therefore, at the brane we have%
\begin{equation}
n_{\mu }\mathrm{d}x^{\mu }=\mathrm{d}\chi ,  \label{br15}
\end{equation}%
and the bulk metric at the brane takes the form, 
\begin{equation}
\mathrm{d}s^{2}=\mathrm{d}\chi ^{2}+\eta _{ab}e^{a}e^{b}.  \label{br16}
\end{equation}

Using the normal coordinate, we separate the stress-energy tensors in the
same way as in Ref.~\cite{br8},%
\begin{eqnarray}
T_{AB} &=&-\Lambda \bar{\eta}_{AB}+\bar{T}_{AB}\delta \left( \chi \right)
\label{tt} \\
T_{ab} &=&-\lambda \eta _{ab}+\bar{\tau}_{ab}  \label{ttt} \\
T_{AB}^{(h)} &=&\bar{T}_{AB}^{(h)}\delta \left( \chi \right) ,
\end{eqnarray}%
where $\lambda $ is the brane tension, $\tau _{ab}$ is the stress-energy
tensor of the matter on the brane, and $T_{ab}^{(h)}$ is the stress-energy
tensor associated with the $h^{a}$ field in the brane. By imposing the
symmetry $Z_{2}$, the juncture condition takes the form

\begin{equation}
K^{k}\,_{d}=-\frac{k_{5}}{2}\left( \tilde{\bar{T}}^{k}\,_{d}-\frac{1}{3}%
\tilde{\bar{T}}\delta ^{k}\,_{d}\right)  \label{Z2}
\end{equation}

On the other side from the contracted Codazzi equation, we find 
\begin{equation}
0=-\frac{k_{5}}{2}\mathcal{\bar{D}}_{s}\bar{\tau}^{s}\,_{d}-\frac{k_{5}}{2}%
\alpha \mathcal{\bar{D}}_{s}\bar{T}^{(h)s}\,_{d},  \label{446}
\end{equation}%
where, as we have previously seen, $\alpha =-\lambda _{1}/\lambda _{3}$.
Since $\lambda _{1}$ and $\lambda _{3}$ are independent, it implies 
\begin{equation}
\mathcal{\bar{D}}_{s}\bar{\tau}^{s}\,_{d}=0,\text{ \ \ }\mathcal{\bar{D}}_{s}%
\bar{T}^{(h)s}\,_{d}=0,
\end{equation}%
that in the language of forms takes the form 
\begin{equation}
\mathrm{\bar{D}}\ast \mathcal{\bar{T}}^{a}=0,\text{ \ \ }\mathrm{\bar{D}}%
\ast \mathcal{\bar{T}}^{(h)a}=0,
\end{equation}%
where%
\begin{eqnarray}
\mathcal{\bar{T}}^{a} &=&\frac{\bar{\tau}}{6}e^{a}-\frac{1}{2}\bar{\tau}%
^{a}\,_{l}e^{l}, \\
\mathcal{\bar{T}}^{a(h)} &=&\frac{\bar{T}^{(h)}}{6}e^{a}-\frac{1}{2}\bar{T}%
^{a(h)}\,_{l}e^{l}.
\end{eqnarray}

Using these results in the equations of motion (\ref{eom1}-\ref{eom3}) for
the brane, we find, 
\begin{eqnarray}
\varepsilon _{abcd}\bar{R}^{ab}e^{c} &=&\varepsilon _{abcd}\left\{ \frac{%
\Lambda _{4}}{3}e^{a}e^{b}-\frac{k_{5}^{2}}{8}\left( \bar{\Pi}^{ab}+\alpha
^{2}\bar{\Pi}^{ab(h)}\right) -16\pi G_{N}\left( \mathcal{\bar{T}}^{a}+\alpha 
\mathcal{\bar{T}}^{a(h)}\right) e^{b}+\right.  \notag \\
&&\left. +\frac{k_{5}^{2}}{4}\alpha \bar{I}^{ab}-E^{a}\,_{m}e^{m}e^{b}\right%
\} e^{c}  \label{ecs4}
\end{eqnarray}

\begin{eqnarray}
2k_{5}i_{4}\bar{T}_{d}^{(h)}=- &&l^{2}\varepsilon _{abcd}\left\{ \left(
E^{a}\,_{m}e^{m}+\frac{k_{5}}{6}\Lambda e^{a}\right) \left( \bar{R}^{bc}+%
\frac{k_{5}^{2}}{36}\lambda ^{2}e^{b}e^{c}-\frac{k_{5}^{2}}{8}\left( \bar{\Pi%
}^{bc}+\right. \right. \right.  \notag \\
&&+\left. \left. \left. \alpha ^{2}\bar{\Pi}^{bc(h)}\right) -16\pi
G_{N}\left( \mathcal{\bar{T}}^{b}+\alpha \mathcal{\bar{T}}^{b(h)}\right)
e^{c}+\frac{k_{5}^{2}}{4}\alpha \bar{I}^{bc}\right) +\bar{D}K^{a}B_{\text{ \
\ \ }l}^{bc}e^{l}\right\}  \notag \\
&&  \label{ecs5}
\end{eqnarray}%
\begin{equation}
l^{2}\varepsilon _{abcd}\left( E^{c}\,_{m}e^{m}+\frac{k_{5}}{6}\Lambda
e^{c}\right) \left( \mathcal{\bar{D}}h^{d}-\frac{k_{5}}{6}\lambda e^{d}-k_{5}%
\mathcal{\bar{T}}^{d}\right) +l^{2}\varepsilon _{abcd}\bar{D}K^{c}\bar{D}%
_{4}h^{d}=0,  \label{ecs6}
\end{equation}%
where,

\begin{eqnarray}
\bar{I}^{ab} &=&2\bar{T}^{a(h)}\,_{l}\bar{\tau}^{b}\,_{p}e^{l}e^{p}-\frac{2}{%
9}\bar{\tau}\bar{T}^{(h)}e^{a}e^{b}-\frac{2}{3}\bar{\tau}^{a}\,_{l}\bar{T}%
^{(h)}e^{l}e^{b}-\frac{2}{3}\bar{\tau}\bar{T}^{a(h)}\,_{l}e^{l}e^{b},  \notag
\\
&&  \label{ecs7} \\
\bar{\Pi}^{ab} &=&\frac{1}{3}\bar{\tau}\bar{\tau}_{\text{ \ }%
l}^{a}e^{l}e^{b}-\frac{1}{2}\bar{\tau}_{\text{ \ }l}^{a}\bar{\tau}_{\text{ \ 
}m}^{b}e^{l}e^{m}-\frac{\bar{\tau}^{2}}{18}e^{a}e^{b},  \label{ecs8} \\
\bar{\Pi}^{ab(h)} &=&\frac{1}{3}\bar{T}^{(h)}\bar{T}_{\text{ \ \ }%
l}^{a(h)}e^{l}e^{b}-\frac{1}{2}\bar{T}_{\text{ \ \ }l}^{a(h)}\bar{T}_{\text{
\ \ }m}^{b(h)}e^{l}e^{m}-\frac{\left( \bar{T}^{(h)}\right) ^{2}}{18}%
e^{a}e^{b}  \label{ecs9}
\end{eqnarray}

In the normal direction the equations take the form%
\begin{eqnarray}
l^{2}\varepsilon _{abcd}B^{ab}\,_{l}R^{cd}e^{l} &=&8k_{5}i_{4}\bar{T}%
^{(h)}\,_{4}-l^{2}\varepsilon _{abcd}\left( B^{ab}\,_{l}e^{l}\left( \frac{%
k_{5}^{2}}{36}\lambda ^{2}e^{c}e^{d}-\frac{k_{5}^{2}}{8}\left( \bar{\Pi}%
^{cd}+\right. \right. \right.  \notag \\
&&\left. +\alpha ^{2}\bar{\Pi}^{cd(h)}\right) -16\pi G_{N}\left( \mathcal{%
\bar{T}}^{c}+\alpha \mathcal{\bar{T}}^{c(h)}\right) e^{d}+  \notag \\
&&\left. \left. +\frac{k_{5}^{2}}{4}\alpha \bar{I}^{cd}\right) \right)
\label{c1}
\end{eqnarray}

\begin{eqnarray}
l^{2}\varepsilon _{abcd}R^{bc}\mathcal{D}_{4}h^{d} &=&-l^{2}\varepsilon
_{abcd}\left( \left( \frac{k_{5}^{2}}{36}\lambda ^{2}e^{c}e^{d}-\frac{%
k_{5}^{2}}{8}\left( \bar{\Pi}^{cd}+\alpha ^{2}\bar{\Pi}^{cd(h)}\right)
-16\pi G_{N}\left( \mathcal{\bar{T}}^{c}+\right. \right. \right.  \notag \\
&&\left. \left. \left. +\alpha \mathcal{\bar{T}}^{c(h)}\right) e^{d}+\frac{%
k_{5}^{2}}{4}\alpha \bar{I}^{cd}\right) \mathcal{D}_{4}h^{d}+B^{bc}\,_{l}%
\mathcal{D}h^{d}e^{l}\right)  \label{c2}
\end{eqnarray}

So far, we have obtained the equations of motion (\ref{ecs4},\ref{ecs5}) y (%
\ref{ecs6}) for the brane, besides the junction conditions (\ref{Z2}), and
conditions (\ref{c1}) and (\ref{c2}). The equation (\ref{ecs4}) contains
terms of order one and quadratics in both types of matter, $\mathcal{T}^{a}$%
, $\Pi ^{ab}$, $\mathcal{T}^{(h)a}$, and $\Pi ^{(h)ab}$, as well as an
interaction term $I^{ab}$ between both types of matter. If the interaction
term, the quadratic terms and the simple terms in $T^{(h)ab}$ are of the
same or smaller order of magnitude, they can be neglected if $\alpha $ is
small.

The system of equations given by (\ref{ecs4},\ref{ecs5}) and (\ref{ecs6})
cannot be solved unless we also consider the equations (\ref{c1}) and (\ref%
{c2}) in addition to the equations describing both parts of the Weyl tensor
in bulk given in the appendix to \cite{br8}.

Taking the exterior covariant derivative of the equation (\ref{ecs4}), and
using the Bianchi identity, we obtain 
\begin{equation}
\varepsilon _{abcd}\mathcal{\bar{D}}E^{a}\,_{m}e^{m}e^{b}e^{c}=\varepsilon
_{abcd}\left\{ -\frac{k_{5}^{2}}{8}\left( \mathcal{\bar{D}}\bar{\Pi}%
^{ab}+\alpha ^{2}\mathcal{\bar{D}}\bar{\Pi}^{ab(h)}\right) +\frac{k_{5}^{2}}{%
4}\alpha \mathcal{D}\bar{I}^{ab}\right\} e^{c}  \label{consE}
\end{equation}

In contrast to the case of Ref.~\cite{br8}, here the electrical part of the
Weyl tensor is also restricted by $T^{a(h)}\,_{l}$ and the interaction $%
I^{ab}$.

Let us compare the new terms in the equations to the stress-energy tensor $%
\tau ^{a}\,_{b}$, in a way similar to Ref.~\cite{br8}. We set the scales of
the constants 
\begin{eqnarray}
k_{5} &=&\frac{1}{M_{G}^{3}},\text{ \ \ \ }\lambda =M_{\lambda }^{4} \\
\left\vert \bar{\tau}_{b}^{a}\right\vert &=&M^{4},\text{ \ \ \ \ }\left\vert 
\bar{\tau}_{b}^{a(h)}\right\vert =M_{(h)}^{4},
\end{eqnarray}%
where $M_{G}$ and $M_{\lambda }$ are larger than the characteristic energy
scales $M$ and $M_{(h)}$, with $M$ being of the same or greater order of
magnitude than $M_{(h)},$

\begin{eqnarray}
k_{5}^{2}\alpha ^{2}\frac{\left\vert \varepsilon _{abcd}\bar{\Pi}%
^{ab(h)}e^{c}\right\vert }{G_{N}\left\vert \varepsilon _{lmns}\mathcal{\bar{T%
}}^{l}e^{m}e^{n}\right\vert } &\sim &\alpha ^{2}\frac{M_{(h)}^{8}}{%
M_{\lambda }^{4}M^{4}}  \label{c11} \\
\alpha \frac{G_{N}\left\vert \varepsilon _{abcd}\mathcal{\bar{T}}%
^{a(h)}e^{b}e^{c}\right\vert }{G_{N}\left\vert \varepsilon _{lmns}\mathcal{%
\bar{T}}^{l}e^{m}e^{n}\right\vert } &\sim &\alpha \frac{M_{(h)}^{4}}{M^{4}}
\\
\alpha \frac{k_{5}^{2}\left\vert \varepsilon _{abcd}\bar{I}%
^{ab}e^{c}\right\vert }{G_{N}\left\vert \varepsilon _{lmns}\mathcal{\bar{T}}%
^{l}e^{m}e^{n}\right\vert } &\sim &\alpha \frac{M_{(h)}^{4}}{M_{\lambda }^{4}%
}.  \label{c12}
\end{eqnarray}%
From the equations (\ref{c11})-(\ref{c12}), we can see, including the case $%
M=M_{(h)}$, that, when $\alpha $ is little, these terms are negligible
compared to the tensor $\tau ^{ab}$ components.

It is useful to separate $E_{ab}$ into two parts, namely, into a transverse
part (no trace), $E_{(TT)}$, and a longitudinal part, $E_{(L)}$, where only
the latter is determined by the matter, since $E_{(TT)}$ corresponds to the
part that interacts between the brane and the bulk.

The longitudinal part of $E_{ab}$ is restricted, as can be seen in the
equation (\ref{consE}), both by the quadratic term in $\tau _{ab}$, and by $%
T^{(h)}$.

Comparing $E_{ab}$ with $\tau _{ab}$ we find

\begin{eqnarray}
\frac{\left\vert \varepsilon _{abcd}E_{(L)m}^{a}e^{m}e^{b}e^{c}\right\vert }{%
G_{N}\left\vert \varepsilon _{lmns}\mathcal{\bar{T}}^{l}e^{m}e^{n}\right%
\vert } &\sim &\frac{1}{G_{N}\left\vert \tau _{ab}\right\vert }\left\vert
G_{N}\alpha T_{ab}^{(h)}+k_{5}^{2}\left( \tau _{al}\tau
^{l}\,_{b}+...\right) +\right.  \notag \\
&&+k_{5}^{2}\alpha ^{2}\left( T_{al(h)}T^{l(h)}\,_{b}+...\right)
+k_{5}^{2}\alpha \left( \tau _{al}T^{l(h)}\,_{b}+...\right) ,  \notag \\
&&
\end{eqnarray}%
where 
\begin{equation*}
\frac{\left\vert \varepsilon _{abcd}E_{(L)m}^{a}e^{m}e^{b}e^{c}\right\vert }{%
G_{N}\left\vert \varepsilon _{lmns}\mathcal{\bar{T}}^{l}e^{m}e^{n}\right%
\vert }\sim \alpha \frac{M_{(h)}^{4}}{M^{4}}+\frac{M^{4}}{M_{\lambda }^{4}}%
+\alpha ^{2}\frac{M_{(h)}^{8}}{M^{4}M_{\lambda }^{4}}+\alpha \frac{%
M_{(h)}^{4}}{M_{\lambda }^{4}},
\end{equation*}%
i.e., the electrical part of the Weyl tensor would be negligible as long as $%
\alpha $ and $l$ are small.

It is interesting to notice that when the $R^{ab}$ components are small,
then the limit $\alpha \longrightarrow 0$ and $l\rightarrow 0$ lead to the
known results of~\cite{br8}. Indeed, the equations of motion take the form,

\begin{eqnarray}
\varepsilon _{abcd}\bar{R}^{ab}e^{c} &=&\varepsilon _{abcd}\left\{ \frac{%
\Lambda _{4}}{3}e^{a}e^{b}-\frac{k_{5}^{2}}{2}\bar{\Pi}^{ab}-16\pi G_{N}%
\mathcal{\bar{T}}^{a}e^{b}-E^{a}\,_{m}e^{m}e^{b}\right\} e^{c},  \notag \\
\varepsilon _{abck}\bar{T}_{\text{ \ \ \ \ }d}^{(h)k}\text{ }e^{a}e^{b}e^{c}
&=&0.  \label{shiro}
\end{eqnarray}%
In this limit equation for $\omega ^{ab}$ is identically null as well as the
tensor $\bar{T}_{\text{ \ \ \ \ }d}^{(h)k}$ . The equation (\ref{shiro})
matches the equation $(17)$ of the reference~\cite{br8}. Note also that in
this limit the equation (\ref{446}) implies the conservation of $\tau _{%
\text{ \ }b}^{a}$.

\section{\textbf{Concluding remarks}}

This article shows that it is possible to obtain both the Lagrangean and the
equations of motions for a 3-brane in 5-dimensional Einstein-Chern-Simons
gravity.

We constructed the Einstein-Chern-Simons gravity juncture conditions
starting from the Lovelock boundary term of Ref.~\cite{gravanis1}, using
AdS-Chern-Simons as an intermediate step. The $S$-expansion procedure closes
the gap between both Chern-Simons theories, mapping one boundary term into
the other. The key is to consider the extra $h^{A}$ field as a matter field.

The new junction condition obtained for the extrinsic curvature corresponds
to the Darmois-Israel joint condition plus a correction, which corresponds
to the matter $T^{(h)}$, which vanishes at the low energy limit.

The procedure described in Ref.~\cite{br8} lead in this case to the
effective equations of motion for a 3-brane embedded in a five-dimensional
space obeying the Einstein-Chern-Simons field equations with $T^{a}=0$ and $%
k_{ab}=0$.

The imposition of the mentioned junction conditions leads to equations for
the brane, showing new terms corresponding to the new type of matter and the
interaction between it and the usual matter $\tau ^{a}\,_{b}$.

These terms disappear in the limit $l\rightarrow 0$, leading to the former
case studied in the reference~\cite{br8}. The cosmological implications of
these new terms will be studied elsewhere (work in progress).

The generalized Poincar\'{e} algebra $\mathfrak{B}_{n}$\textbf{\ }\cite{seba}%
\textbf{, }\cite{salg1}\textbf{, }\cite{concha}, can be obtained from the
anti-de-Sitter algebra and the semigroup $S_{E}^{2n-1}=\left\{ \lambda
_{0},\cdot \cdot \cdot ,\lambda _{2n}\right\} $ whose multiplication law is
given by $\lambda _{\alpha }\lambda _{\beta }=\lambda _{\alpha +\beta }$
when $\alpha +\beta \leq 2n$ and $\lambda _{\alpha }\lambda _{\beta
}=\lambda _{2n}$ when $\alpha +\beta >2n$, where $\lambda _{2n}$ corresponds
to the zero element of the semigroup. The generators of $\mathfrak{B}_{n}$
denoted by $\left( P_{a},J_{ab},Z_{ab}^{(i)},Z_{a}^{(i)}\right) $ satisfy
the following commutation relations \textbf{\ }%
\begin{align}
\left[ P_{a},P_{b}\right] & =\Lambda Z_{ab}^{\left( 1\right) },\text{ \ \ }%
\left[ J_{ab},P_{c}\right] =\eta _{bc}P_{a}-\eta _{ac}P_{b},  \notag \\
\left[ J_{ab},J_{cd}\right] & =\eta _{bc}J_{ad}+\eta _{ad}J_{bc}-\eta
_{ac}J_{bd}-\eta _{bd}J_{ac},  \notag \\
\left[ J_{ab},Z_{c}^{(i)}\right] & =\eta _{bc}Z_{a}^{(i)}-\eta
_{ac}Z_{b}^{(i)},  \notag \\
\left[ Z_{ab}^{\left( i\right) },P_{c}\right] & =\eta _{bc}Z_{a}^{(i)}-\eta
_{ac}Z_{b}^{(i)},\text{ }  \notag \\
\left[ Z_{ab}^{\left( i\right) },Z_{c}^{(j)}\right] & =\eta
_{bc}Z_{a}^{(i+j)}-\eta _{ac}Z_{b}^{(i+j)},  \notag \\
\left[ J_{ab},Z_{cd}^{\left( i\right) }\right] & =\eta _{bc}Z_{ad}^{\left(
i\right) }+\eta _{ad}Z_{bc}^{\left( i\right) }-\eta _{ac}Z_{bd}^{\left(
i\right) }-\eta _{bd}Z_{ac}^{\left( i\right) },  \notag \\
\left[ Z_{ab}^{\left( i\right) },Z_{cd}^{\left( j\right) }\right] & =\eta
_{bc}Z_{ad}^{\left( i+j\right) }+\eta _{ad}Z_{bc}^{\left( i+j\right) }-\eta
_{ac}Z_{bd}^{\left( i+j\right) }-\eta _{bd}Z_{ac}^{\left( i+j\right) }, 
\notag \\
\text{\ }\left[ P_{a},Z_{b}^{(i)}\right] & =Z_{ab}^{\left( i+1\right) }, 
\notag \\
\left[ Z_{a}^{(i)},Z_{b}^{(j)}\right] & =Z_{ab}^{\left( i+j+1\right) }
\label{ej5}
\end{align}%
where, $\tilde{J}_{ab}$\ and $\tilde{P}_{a}$\ are the generators of the
anti-de-Sitter algebra and $J_{ab}=\lambda _{0}\otimes \tilde{J}_{ab},$\ $%
Z_{ab}^{\left( i\right) }=\lambda _{2i}\otimes \tilde{J}_{ab},$\ $%
P_{a}=\lambda _{1}\otimes \tilde{P}_{a}$\ and $Z_{a}^{(i)}=\lambda
_{2i+1}\otimes \tilde{P}_{a}$, \ with $i,j=0,1,\cdot \cdot \cdot ,n-1$\ are
the generators of the $B_{n}$ algebra.\ \ 

This means that the results obtained so far can be generalized to the case
of generalized Poincare algebras $\mathfrak{B}_{n}$ where $n$ can be either
even $\left( \mathfrak{B}_{2m}\right) $ or odd $\left( \mathfrak{B}%
_{2m+1}\right) $ and subsequently study their respective applications in
black holes and in cosmology. Works in this direction are in progress.

\section{\textbf{Appendix 1: Gauss-Codazzi equations in the Cartan formalism 
}}

Let us consider an $n$-dimensional manifold $\Sigma $ immersed into an $m$%
-dimensional manifold $M$.

For the $m$-dimensional manifold $M$, at each point $P$ we define a
cotangent space $T_{P}^{\ast }(M)$, and a local coordinate system $y^{\mu }$%
. It allows us to define a coordinate base $\left\{ \mathrm{d}x^{\mu
}\right\} _{\mu =1}^{\mu =m}$, with $\mu ,\nu ,\cdots =1,\cdots ,m$, and an
orthonormal basis $e^{A}=e^{A}{}_{\mu }\mathrm{d}x^{\mu }$, with $A,B,\cdots
=1,\cdots ,m$, so that $e^{A}\cdot e^{B}=\eta ^{AB}$ and and $g^{\mu \nu }=%
\mathrm{d}x^{\mu }\cdot \mathrm{d}x^{\nu }$. Similarly, for the $n$%
-dimensional manifold, at each point $P$ we define a cotangent space $%
T_{P}^{\ast }(\Sigma )$ and a local coordinate system $x^{i}$. It allows us
to define a coordinate base $\left\{ \mathrm{d}x^{i}\right\} _{i=1}^{i=n},$ $%
i,j,\cdots =1,\cdots ,n$, and an orthonormal basis $\bar{e}^{a}=\bar{e}%
^{a}{}_{i}\mathrm{d}x^{i}$, with $a,b,\cdots =1,\cdots ,n$, such that $\bar{e%
}^{a}\cdot \bar{e}^{b}=\eta ^{ab}$ and $g^{ij}=\mathrm{d}x^{i}\cdot \mathrm{d%
}x^{j}$. We will use the indices $r,s,\cdots =n+1=m$ for a base $M$
orthogonal to $\Sigma $.

Let us consider the application of the structure equations to the spaces $%
\Sigma $ and $M$ with the null torsion condition. We have $e^{a}=\bar{e}^{a}$%
, $e^{n+1}=0$, and $\omega ^{a}{}_{c}=\bar{\omega}^{a}{}_{c}$ when
pullbacked to $T_{P}^{\ast }(\Sigma )$ From the first equation of structure $%
T^{\text{ }A}=\mathrm{d}e^{A}+\omega _{\text{ }C}^{A}e^{C}=0$, we have that $%
\bar{T}^{\text{ }a}=\mathrm{d}\bar{e}^{a}+\bar{\omega}^{a}{}_{c}\bar{e}%
^{c}=0 $ and $\omega ^{n+1}\,_{c}\wedge e^{c}=0$ . From the Cartan lemma we
can write that $\omega ^{n+1}\,_{c}=K_{ac}e^{a}$, that is, $K_{ac}=\omega
^{n+1}\,_{a}\cdot e_{c}$ which corresponds to the Gaussian-Weingarten
equations, where $K_{ac}$ is the extrinsic curvature.

From the second structure equation over $M,$ $\ R_{\text{ }B}^{A}=\mathrm{d}%
\omega _{\text{ }B}^{A}\,+\omega _{\text{ }C}^{A}\omega _{\text{ }B}^{C}$ \
it is straightforward to see that for the indices $a,b$ we have $R_{\text{ }%
b}^{a}=\bar{R}_{\text{ }b}^{a}+K_{\text{ }n}^{a}K_{\text{ }bm}e^{n}e^{m}$
and for the indices $n+1,b$ we have $R_{\text{ }b}^{n+1}=\left( dK_{\text{ }%
bf}^{(n+1)}\,-K_{\text{ }cf}^{(n+1)}\omega _{\text{ }b}^{c}-K_{\text{ }%
sf}^{(n+1)}\omega _{\text{ }b}^{s}\right) e^{f}$, which corresponds to the
Codazzi equation.

\section{\textbf{Appendix 2: Israel Junction Conditions }}

The presence of a hypersurface $\Sigma $, in a manifold $M$ divides
space-time into two regions $M^{+}$ and $M^{-}$ that have $\Sigma $ as
boundary, and $g_{\alpha \beta }^{+}$ and $g_{\alpha \beta }^{-}$
respectively as metrics. Let us call $\xi ^{a},$ $a=1,2,3$ to the
\textquotedblleft intrinsic\textquotedblright \ coordinates on both faces of
the hypersurface $\Sigma $, and $x_{\pm }^{\alpha }$, $\alpha =0,1,2,3$ to
the coordinates of the varieties $M^{\pm }$.

Let us define a normal vector $N_{\alpha }=\varepsilon \partial _{\alpha
}\ell $ to the hypersurface, such that $N^{\alpha }N_{\alpha }=\varepsilon $
and they point from $M^{-}$ to $M^{+}$ \cite{poisson}, where $\ell $ denotes
the proper distance along the geodesics, so that $\ell =0$ when the
geodesics traverses the hypersurface.

The step function $\Theta \left( \ell \right) $ is defined, equal to $+1$ if 
$\ell >0$, 0 if $\ell <0$, such that 
\begin{equation*}
\Theta ^{2}\left( \ell \right) =\Theta \left( \ell \right) \text{, \ \ }%
\Theta \left( \ell \right) \Theta \left( -\ell \right) =0\text{, \ }\frac{%
\mathrm{d}}{\mathrm{d}\ell }\Theta \left( \ell \right) =\delta \left(
l\right) ,
\end{equation*}%
where $\delta \left( l\right) $ is the Dirac distribution. We will denote
with the symbol $\left[ {}\right] $ the "jump" of a tensor quantity $\Omega $
through the hypersurface $\Sigma $

\begin{equation*}
\left[ \Omega \right] \equiv \left. \Omega \left( M^{+}\right) \right \vert
_{\Sigma }-\left. \Omega \left( M^{-}\right) \right \vert _{\Sigma }\text{,}
\end{equation*}%
where $\Omega $ is defined on both sides of the hypersurface. The equation $%
\left[ N^{\alpha }\right] =0$ follows from the relation $N_{\alpha
}=\varepsilon \partial _{\alpha }\ell $ and the continuity of both $\ell $
and $x^{\alpha }$ through $\Sigma $. The equation $\left[ X_{a}^{\alpha }%
\right] =0$ follows from the fact that the coordinates $\xi ^{a}$ are the
same on both sides of the hypersurface \cite{poisson}.

We can write the metric $g_{\alpha \beta }$ as

\begin{equation}
g_{\alpha \beta }=\Theta \left( \ell \right) g_{\alpha \beta }^{+}+\Theta
\left( -\ell \right) g_{\alpha \beta }^{-}  \label{isr13'}
\end{equation}%
where $g_{\alpha \beta }^{\pm }$ is the metric in $M^{\pm }$ expressed in
the coordinates $x^{\alpha }$. From here it is direct to see that

\begin{equation}
g_{\alpha \beta /\gamma }=\Theta \left( \ell \right) g_{\alpha \beta /\gamma
}^{+}+\Theta \left( -\ell \right) g_{\alpha \beta /\gamma }^{-}+\varepsilon
\delta \left( l\right) \left[ g_{\alpha \beta }\right] N_{\gamma },
\label{irs13''}
\end{equation}%
where the last term is singular. To solve this problem we must impose the
continuity of the metric through the hypersurface, $\left[ g_{\alpha \beta }%
\right] =0$, a condition that can be rewritten in the form 
\begin{equation*}
\left[ g_{\alpha \beta }\right] X_{a}^{\alpha }X_{b}^{\beta }=\left[
g_{\alpha \beta }X_{a}^{\alpha }X_{b}^{\beta }\right] =\left[ \gamma _{ab}%
\right] =0
\end{equation*}%
and that is known as the first joint condition. A direct calculation shows
that the Riemann tensor is given by

\begin{equation}
R_{\text{ \ }\beta \gamma \delta }^{\alpha }=\Theta \left( \ell \right) R_{%
\text{ \ }\beta \gamma \delta }^{+\alpha }+\Theta \left( -\ell \right) R_{%
\text{ \ }\beta \gamma \delta }^{-\alpha }+\delta \left( l\right) A_{\text{
\ }\beta \gamma \delta }^{\alpha },  \label{irs13v}
\end{equation}%
where%
\begin{align}
R_{\text{ \ }\beta \gamma \delta }^{+\alpha }& =\Gamma _{\text{ \ }\beta
\delta /\gamma }^{+\alpha }-\Gamma _{\text{ \ }\beta \gamma /\delta
}^{+\alpha }+\Gamma _{\text{ \ \ }\mu \gamma }^{+\alpha }\Gamma _{\text{ \ }%
\beta \delta }^{+\mu }-\Gamma _{\text{ \ }\mu \delta }^{+\alpha }\Gamma _{%
\text{ \ \ }\beta \gamma }^{+\mu }  \notag \\
R_{\text{ \ }\beta \gamma \delta }^{-\alpha }& =\Gamma _{\text{ \ }\beta
\delta /\gamma }^{-\alpha }-\Gamma _{\text{ \ }\beta \gamma /\delta
}^{-\alpha }+\Gamma _{\text{ \ \ }\mu \gamma }^{-\alpha }\Gamma _{\text{ }%
\beta \delta }^{-\mu }-\Gamma _{\text{ \ \ }\mu \delta }^{-\alpha }\Gamma _{%
\text{ \ \ }\beta \gamma }^{-\mu }  \notag \\
A_{\text{ \ }\beta \gamma \delta }^{\alpha }& =\varepsilon \left( \left[
\Gamma _{\text{ }\beta \delta }^{\alpha }\right] N_{\gamma }-\left[ \Gamma _{%
\text{ }\beta \gamma }^{\alpha }\right] N_{\delta }\right) .  \label{irsvi}
\end{align}

An explicit expression for the tensor $A_{\text{ \ }\beta \gamma \delta
}^{\alpha }$ can be obtained taking into account that the metric is
continuous through $\Sigma $. If $g_{\alpha \beta /\gamma }$ were
discontinuous, then the discontinuity must be directed along the normal
vector $N^{\alpha }$, which implies that there must exist a field $\kappa
_{\alpha \beta }$ such that $\left[ g_{\alpha \beta /\gamma }\right] =\kappa
_{\alpha \beta }N_{\gamma }$, and therefore

\begin{equation*}
\left[ \Gamma _{\text{ }\beta \gamma }^{\alpha }\right] =\frac{1}{2}\left(
\kappa _{\text{ }\beta }^{\alpha }N_{\gamma }+\kappa _{\text{ \ }\gamma
}^{\alpha }N_{\beta }-\kappa _{\beta \gamma }N^{\alpha }\right) .
\end{equation*}

This implies that

\begin{eqnarray}
A_{\alpha \beta } &\equiv &A_{\text{ \ }\alpha \mu \beta }^{\mu }=\frac{%
\varepsilon }{2}\left( \kappa _{\mu \alpha }N^{\mu }N_{\beta }+\kappa _{\mu
\beta }N^{\mu }N_{\alpha }-\kappa N_{\alpha }N_{\beta }-\varepsilon \kappa
_{\alpha \beta }\right) ,  \label{irs13-4} \\
A &\equiv &A_{\text{ }\alpha }^{\alpha }=\varepsilon \left( \kappa _{\mu \nu
}N^{\mu }N^{\nu }-\varepsilon \kappa \right) .  \label{irs13-5}
\end{eqnarray}%
where $\kappa =\kappa _{\text{ \ }\mu }^{\mu }$.

The stress-energy tensor corresponds to~\cite{poisson} 
\begin{equation}
T_{\alpha \beta }=\Theta \left( \ell \right) T_{\alpha \beta }^{+}+\Theta
\left( -\ell \right) T_{\alpha \beta }^{-}+\delta \left( \ell \right)
S_{\alpha \beta },  \label{irs13-6}
\end{equation}%
where $T_{\alpha \beta }^{+}$ and $T_{\alpha \beta }^{-}$ are the
stress-energy tensors of the $M^{+}$ and $M^{-}$ regions, and $S_{\alpha
\beta }$ is the stress-energy tensor associated with $\Sigma $.

In the Einstein's field equations

\begin{equation*}
G_{\alpha \beta }=\kappa T_{\alpha \beta },
\end{equation*}%
we have 
\begin{equation}
G_{\alpha \beta }=\Theta \left( \ell \right) G_{\alpha \beta }^{+}+\Theta
\left( -\ell \right) G_{\alpha \beta }^{-}+\delta \left( \ell \right)
G_{\alpha \beta }^{\Sigma },  \label{irs13-7}
\end{equation}%
and $T_{\alpha \beta }$ corresponds to~(\ref{irs13-6}). Therefore%
\begin{align*}
G_{\alpha \beta }^{+}& =R_{\alpha \beta }^{+}-\frac{1}{2}g_{\alpha \beta
}^{+}R^{+}=\kappa T_{\alpha \beta }^{+} \\
G_{\alpha \beta }^{-}& =R_{\alpha \beta }^{+}-\frac{1}{2}g_{\alpha \beta
}^{+}R^{+}=\kappa T_{\alpha \beta }^{-} \\
G_{\alpha \beta }^{\Sigma }& =A_{\alpha \beta }-\frac{1}{2}g_{\alpha \beta
}A=\kappa S_{\alpha \beta },
\end{align*}%
where $S_{ab}=S_{\alpha \beta }X_{a}^{\alpha }X_{b}^{\beta }$ \ takes the
form

\begin{eqnarray}
S_{ab} &=&-\frac{\varepsilon }{\kappa }\left( \left[ K_{ab}\right] -\gamma
_{ab}\left[ K\right] \right) ,  \label{irs13-13'} \\
\left[ K_{ab}\right] &=&-\varepsilon \kappa \left( S_{ab}-\frac{1}{2}\gamma
_{ab}S\right) ,  \label{irs13-14}
\end{eqnarray}%
equation known as the Lanczos equation~\cite{poisson}.

\begin{acknowledgement}
This work was supported in part by\textit{\ }FONDECYT Grants\textit{\ }No.%
\textit{\ }1180681 and No 1211219 from the Government of Chile. One of the
authors (RS) was supported by Universidad de Concepci\'{o}n, Chile.
\end{acknowledgement}

\end{document}